\documentclass[showpacs,prb,reprint,floatfix,superscriptsaddress]{revtex4-2}
\usepackage{graphicx}
\usepackage{amssymb}
\usepackage{amsmath}
\usepackage[utf8]{inputenc}
\usepackage{bm}
\usepackage{braket}
\usepackage{hyperref}
\usepackage{color}
\usepackage{float}

\newcommand{\down}{\downarrow}
\newcommand{\up}{\uparrow}

\begin{document}
\title{Trimers in the Extended Hubbard Model}

\author{R. R. Montenegro-Filho}
\affiliation{Laborat\'{o}rio de F\'{i}sica Te\'{o}rica e Computacional, Departamento de F\'{i}sica, Universidade Federal de Pernambuco, 50760-901 Recife-PE, Brasil}
\author{D. R. B. Silva}
\affiliation{Laborat\'{o}rio de F\'{i}sica Te\'{o}rica e Computacional, Departamento de F\'{i}sica, Universidade Federal de Pernambuco, 50760-901 Recife-PE, Brasil}
\author{D. Cogollo}
\affiliation{Departamento de F\'{i}sica, Universidade Federal de Campina Grande, Campina Grande, PB, Brazil}
\author{M. D. Coutinho-Filho}
\affiliation{Laborat\'{o}rio de F\'{i}sica Te\'{o}rica e Computacional, Departamento de F\'{i}sica, Universidade Federal de Pernambuco, 50760-901 Recife-PE, Brasil}
\date{\today}

\begin{abstract}
The Lieb theorem is a cornerstone of quantum magnetism theory in condensed matter. In this work, we investigate the instability of the Lieb insulating ferrimagnetic phase in the extended Hubbard model on a trimer chain at half-filling, with one electron per site, under increasing the nearest-neighbor Coulomb coupling $V$. Our results show that despite a noticeable increase in doublon density with $V$, the ferrimagnetic insulating phase remains robust up to the phase separation (PS) line, which is observed at $V \gtrsim U/4$, where $U$ is the local Coulomb repulsion. Above the PS line, one of the coexisting phases is primarily populated by doublons on one of the two sublattices of the chain. This phase coexists with a metallic, unsaturated ferromagnetic phase for $U \gtrsim t$, and with a singlet phase for $U \lesssim t$, where $t$ is the intra-trimer hopping amplitude. We estimate the PS and the crossover lines with the help of density matrix renormalization group calculations.

\end{abstract}

\maketitle

\section{Introduction}

It is widely recognized that the role played by the originally proposed three-dimensional Hubbard model \cite{hubbard1963electron}, with a focus on electronic phenomena of transition metals and alloys, has gone far beyond early expectations \cite{moriya,Arovas2021}. In fact, the use of this model in the description of a variety of electronic phenomena involving both charge and spin excitations in low-dimensional condensed matter systems is, indeed,  superlative \cite{Arovas2021}. The investigation of the Hubbard model on a two-dimensional square lattice, particularly with the aid of numerical methods \cite{qin2022}, has been fundamental in advancing the understanding of the phase diagram of high-temperature cuprate superconductors \cite{zheng2017,xu2022,simkovic2024}.

While some aspects of the two-dimensional version remain elusive \cite{Arovas2021}, the Lieb-Wu solution \cite{lieb1968absence} of the one-dimensional version \cite{ha1996quantum} has motivated a great amount of theoretical and experimental findings \cite{essler}, including the exotic ones associated with the Luttinger liquid behavior \cite{giamarchi2004quantum}. In this context, the extended version of the one-dimensional Hubbard model has been proposed \cite{*[{See, e.g., }][{, and references therein.}] hirsch1984} and the occurrence and description \cite{nakamura2000,sengupta2002, jeckelmann2002, tsuchiizu2002,sandvik2004,zhang2004,ejima2007} of charge density wave (CDW), spin density wave (SDW), and bond charge density wave (BCDW) have been identified \cite{voit1992} at strong \cite{vandongen1994} and weak coupling \cite{vandongen1994a}, and associated with materials \cite{Arovas2021}. In addition, several interesting features were observed in the one-dimensional extended Hubbard model in the atomic limit \cite{rojas2024} and in a version of the extended Hubbard model in the diamond chain \cite{rojas2021}.

In this work, we present several relevant properties of the referred extended Hubbard model on coupled trimer chains. In particular, very interesting edge states \cite{PhysRevA.99.013833,verma2024} in the tight-binding regime of trimer chains have been revealed. These systems have connections with recent investigations on composite photonic lattices with a broad channel to sustain topological interface states \cite{zhao_interface_2024} and optical coupling functionalities in super-Su-Schrieffer-Heeger lattices \cite{hirsch1984,heeger1988,zhao_interface_2024}. In addition, a recent proposal concerns an interdisciplinary approach to studying topological systems in real space through the combination of information entropy and topological photonics \cite{ma2024}. 

The Heisenberg Hamiltonian (the strong-coupling limit of the Hubbard model at half-filling) on coupled trimer chains exhibits a 1/3 plateau in the magnetization curve as a function of the magnetic field. However, the ground state and specific quantum states depend on how the trimers are coupled. In the paramagnet Cu$_3$(P$_2$O$_6$OD)$_2$ \cite{Hase2007,Hase2020}, there is only one superexchange coupling between neighboring trimers \cite{Gu2006,Hase2007,Hase2020,Cheng2022,Verkholyak2021}, whereas the azurite compound Cu$_3$(CO$_3$)$_2$(OH)$_2$ \cite{Kikuchi2005,Rule2008,Aimo2009} is a nonbipartite diamond chain \cite{Takano1996,Okamoto2003}. In the case of the phosphates A$_3$Cu$_3$(PO$_4$)$_4$ ($A =$ Ca, Sr, Pb) \cite{Matsuda2005,Belik2005}, the superexchange pathways and values can lead to a ferrimagnetic ground state in certain cases \cite{montenegro-filho2022}. Notably, the bipartite $AB_2$ Hubbard model \cite{PhysRevLett.74.1851} displays a rich phase diagram as a function of Coulomb coupling and hole doping away from half-filling, featuring ferrimagnetic insulating and metallic phases, Nagaoka saturated ferromagnetism, Luttinger liquid phases, and phase separation \cite{PhysRevLett.74.1851, PhysA2005, Montenegro-Filho2006, oliveira2009b, Montenegro-Filho2014, MartinezCoutinhJPCM2019}. In particular, the ground-state total spin can be predicted by the Lieb-Mattis theorem \cite{LiebMattis} for the Heisenberg model with only antiferromagnetic superexchange couplings and by Lieb's theorem \cite{lieb1989} for the Hubbard model at half-filling, both on bipartite lattices.

In this study, we investigate the ground state of the extended Hubbard model on a bipartite coupled trimer chain at half-filling, i.e., one particle per site. The model consistently exhibits a Lieb-type quantum ferrimagnetic ground state for any finite value of the on-site Coulomb repulsion 
$U$ and zero nearest-neighbor Coulomb interaction $V$. Using the density matrix renormalization group (DMRG) \cite{White1992,*White1993,*Schollwock2005}, we examine the evolution of this Lieb ferrimagnetic state as $V$ increases and its eventual instability to phase-separated states.

This work is structured as follows. In Sec. II, we introduce the extended Hubbard model Hamiltonian for coupled trimer chains defined by the intra- and intertrimer hopping amplitudes $t$ and $t^\prime$, along with the on-site Coulomb repulsion $U$ and the nearest-neighbor interaction $V$. We also discuss the atomic limit of the model (with $t=0$ and $t^\prime=0$), focusing on the key relevant states and the ferrimagnetic phase of the standard Hubbard model ($U\neq 0$ and $V = 0$). In Sec. III, we analyze the full model, fixing $t \equiv 1$ and $t^\prime = 0.8$. To differentiate the observed quantum phases, we examined spatial averages and local quantum distributions of both the particle number and doublon occupancy within the trimer, as well as the total spin. Finally, in Sec. IV, we provide a summary of our findings and conclusions.

\section{Extended Hubbard Model on coupled trimer chains}
\begin{figure}
\centerline{\includegraphics*[width=0.37\textwidth]{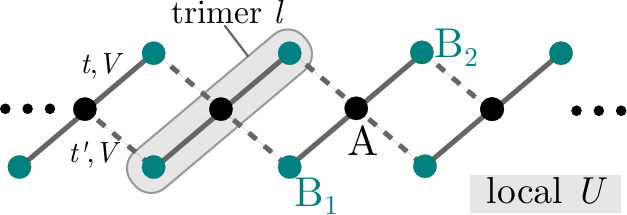}}
\caption{Illustration of the extended Hubbard model for the coupled 
trimer chain. The three sites of a trimer $l$ are identified by $B_1$, $A$, and $B_2$. The intratrimer hopping $t\equiv 1$ (full lines), while $t^\prime$ is the intertrimer 
hopping (dashed lines). The local Coulomb repulsion is $U$ and the nearest-neighbor repulsion is $V$.}
\label{fig:model}
\end{figure}

The extended Hubbard model used in this work is defined 
by the following Hamiltonian:
\begin{align}
 H &= -t\sum_{l=1}^L\sum_{\sigma \in \{\downarrow,\uparrow\}}\left(A^\dagger_{l,\sigma}B_{1l,\sigma}+A^\dagger_{l\sigma}B_{2l,\sigma} + \text{H.c.}\right)\nonumber\\
 &~~-t^\prime \sum_{l=1}^L\sum_{\sigma \in \{\downarrow,\uparrow\}}\left(A^\dagger_{l,\sigma}B_{1,l+1,\sigma}+B^\dagger_{2l,\sigma}A_{l+1,\sigma}+\text{H.c.}\right)\nonumber\\
 &~~+U\sum_i^N n_{i\uparrow}n_{i\downarrow}+V\sum_{<i,j>}^Nn_in_j
 \label{eq:ham}
\end{align}
where $A_{l,\sigma}$ ($A^\dagger_{l,\sigma}$), $B_{1l,\sigma}$ ($B^\dagger_{1l,\sigma}$), and $B_{2l,\sigma}$ ($B^\dagger_{2l,\sigma}$) annihilate (create) electrons of spin $\sigma=\up,\down$ at sites $A$, $B_1$, and $B_2$, respectively, of the 
trimer $l$, see Fig. \ref{fig:model}. There are two hopping terms: an intratrimer, $t$, and an intertrimer, $t^\prime$. Unless 
otherwise noticed, we consider $t\equiv 1$ and $t^\prime=0.8$. $U$ is the local Coulomb repulsion, $V$ is the nearest-neighbor Coulomb repulsion,
$n_{i\sigma}$ are number operators, and $n_i=n_{i\uparrow}+n_{i\downarrow}$. The total number of trimers is $L$,
$i$ runs over all sites, $\langle i,j\rangle$ denotes nearest-neighbor sites, and $N=3L$ is the total number of sites.
In this work, we study the coupled trimer model for one electron per site. We have chosen distinct values for the inter- and intratrimer couplings to avoid the local symmetry \cite{Montenegro-Filho2006} from the interchange of $B_1$ and $B_2$ sites at neighboring trimers, which could lock the renormalization into local minima during the early stages of the procedure.

Crucial quantities to distinguish the observed quantum phases are the average particle number, $\langle n_i\rangle$, and the doublon occupancy $\langle d_i\rangle$ at a given site $i$:
\begin{equation}
\langle d_i\rangle \equiv  \langle n_{i\uparrow}n_{i\downarrow}\rangle.
\label{eq:average-doublon}
\end{equation} 
The spatial averages over all $A$, or $B_1$, or $B_2$ sites are denoted by
$\overline{\braket{n_A}}$, $\overline{\braket{n_{B_1}}}$, $\overline{\braket{n_{B_2}}}$, respectively, in the case of particle occupancy,
and similarly for the doublon occupancy.


\begin{figure}
\centerline{\includegraphics*[width=0.37\textwidth]{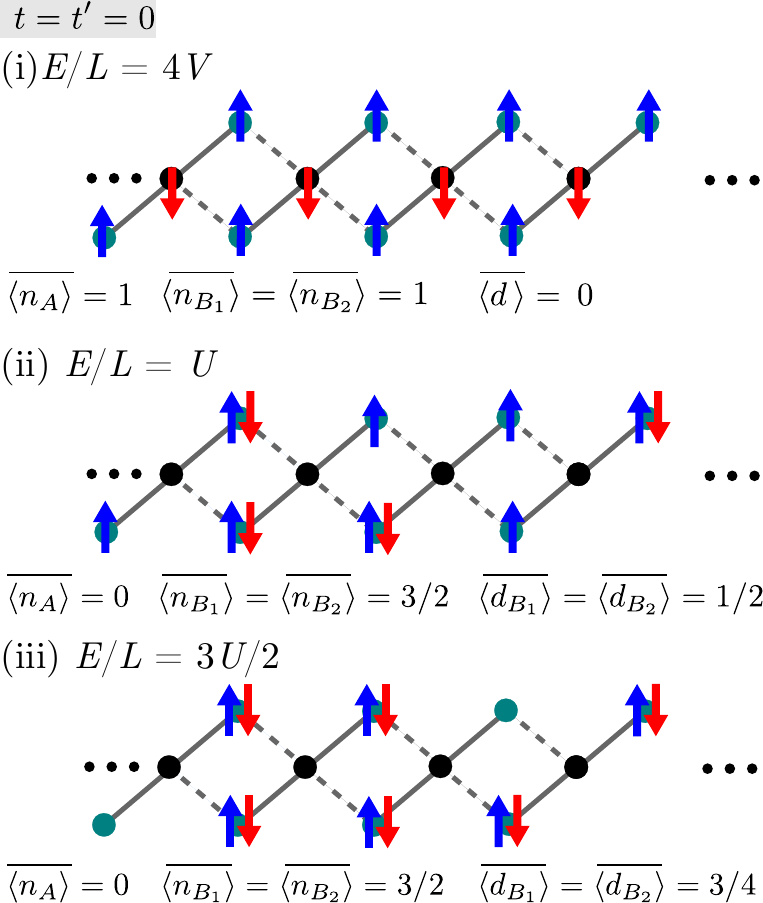}}
\caption{Static configurations for $t=t^\prime=0$, where $E/L$ is the total energy per trimer. 
(i) Static Lieb ferrimagnetic order with average doublon density $\langle d\rangle =0$ and magnetization 
per trimer $m=1/2$, 
(ii) configuration with spatial average density at $A$ sites $\overline{\langle n_A\rangle} =0$, and spatial average doublon density 
at $B$ sites $\overline{\langle d_{B_1}\rangle}=\overline{\langle d_{B_2}\rangle}=1/2$, and $m=1/2$, (iii) the $A$
sites are empty and $\overline{\langle d_{B_1}\rangle}=\overline{\langle d_{B_2}\rangle}=3/4$.}
\label{fig:static-conf}
\end{figure}

\subsection{Standard Hubbard model: $t\neq 0$, $t^\prime\neq 0$, $U\neq0$, and $V=0$}

The ground-state total spin, $S$, of the standard Hubbard model is predicted by Lieb's theorem \cite{lieb1989}, which asserts that for one electron per site (half-filled band) and bipartite lattices:
\begin{equation}
S=|N_B-N_A|/2, 
\end{equation}
where $N_A$ ($N_B$) is the total number of sites in the $A$ ($B$) sublattice. For Hamiltonian (\ref{eq:ham}), $N_A=L$ and $N_B=2L$, thus $S=L/2$, i.e., $1/2$ per trimer. Moreover, the system exhibits an insulating ferrimagnetic long-range ordered phase \cite{Tian1996}, with magnetic moments arranged in the pattern shown in Fig. \ref{fig:static-conf}(i) for the total spin component $S^z=S$.

\subsection{Atomic limit: $t= 0$, $t^\prime= 0$, $U\neq0$, and $V\neq0$}

In Fig. \ref{fig:static-conf} we present some relevant static configurations, corresponding to $t=t^\prime=0$, for 
$U\neq0$ and $V\neq0$. In (i), there is one electron per site, so $\braket{d}=0$ for any site.
Thus, the local Coulomb repulsion $U$ does not contribute to the energy, 
$E/L=4V$, due to the four nearest neighbors of each $A$ site. In (ii),
the $A$ sites are empty, one $B$ site is doubly occupied, while the other $B$ site of the trimer is singly occupied. 
Thus, the nearest-neighbor repulsion is absent, and the energy is given by $E/L = U$. Similarly to configuration (ii), 
state (iii) has also empty $A$ sites, but the number of doubly occupied sites is maximum: $N/2=3L/2$, such that 
$E/L=3U/2$. We remark that the states in Fig. \ref{fig:static-conf} are highly degenerate, both in the charge and spin 
sectors, so they represent a large class of states. In fact, they 
give a good description of the phase transition that might occur in the model with hopping. 

Considering the static configurations, a comparison between the energies of states (i) and (ii) in Fig. \ref{fig:static-conf} shows that if we increase $V$ from $V=0$, a transition to state (ii) occurs 
at 
\begin{equation}
V=U/4. 
\label{eq:trans}
\end{equation}
On the other hand, states (ii) and (iii) become degenerate at $U=0$ and have lower energies 
compared to state (i), in which case $V\neq 0$.

\section{Extended Hubbard model}

In this section, we discuss the phase diagram for distinct
regimes of the model Hamiltonian (\ref{eq:ham}), considering that the intratrimer
hopping is fixed at $t\equiv 1$ and the intertrimer hopping at $t^\prime=0.8$.
The average doublon occupancy, Eq. (\ref{eq:average-doublon}),
as well as the particle occupancy, $\braket{n_i}$, can be used to
assess the phase transitions induced by the nearest-neighbor Coulomb repulsion $V$.

In the calculations, we use the DMRG code from the ITensor library \cite{itensor} in trimer
chains with open boundary conditions and considering
maximum bond dimensions from 512 to 2048, with a maximum truncation error of $\sim 1 \times 10^{-6}$.

{  As shown below, in the case of the extended trimer Hubbard model, the nearest-neighbor coupling $V$ induces phase separation
for a specific value of $V$ which depends on $U$. In the unstable coexistence region, there are two competing phases, with
a distinct average number of particles per trimer. In a finite size system, there is a huge number of competing states
with variable-size clusters of one of the phases embedded in the other in distinct positions along the chain. All these
states have close energies and the renormalization procedure can find one of these states. Apart from these discrepancies,
the states share the physical feature of being inhomogeneous, with two specific values of average number of
particles per trimer in the bulk of the two coexisting phases.

Considering these characteristics of the model, we have performed two kinds of renormalization procedure. In the majority
of the data presented below, we start the simulation from a distribution of one particle per site, with a random distribution
of values of $S_i^z={\uparrow,\downarrow}$, satisfying $\sum_{i}S_i^z=0$. On the other hand, the charge profiles
exhibited in Figs. \ref{fig:chargedistU0} and \ref{fig:localdistUfinitep2} were obtained from an initial charge
distribution with all the $A$ sites empty, and doubly occupied $B$ sites on the left $3L/4$ part of the chain, with the other $B$ sites empty. We confirmed that, in these cases, the energies obtained from the second procedure are lower or nearly equal to the ones obtained from the first methodology.
}

\subsection{$t-t^\prime-V$ model: $t\neq 0$, $t^\prime\neq 0$, $U=0$, and $V\neq0$}
\begin{figure}
\centerline{\includegraphics*[width=0.47\textwidth]{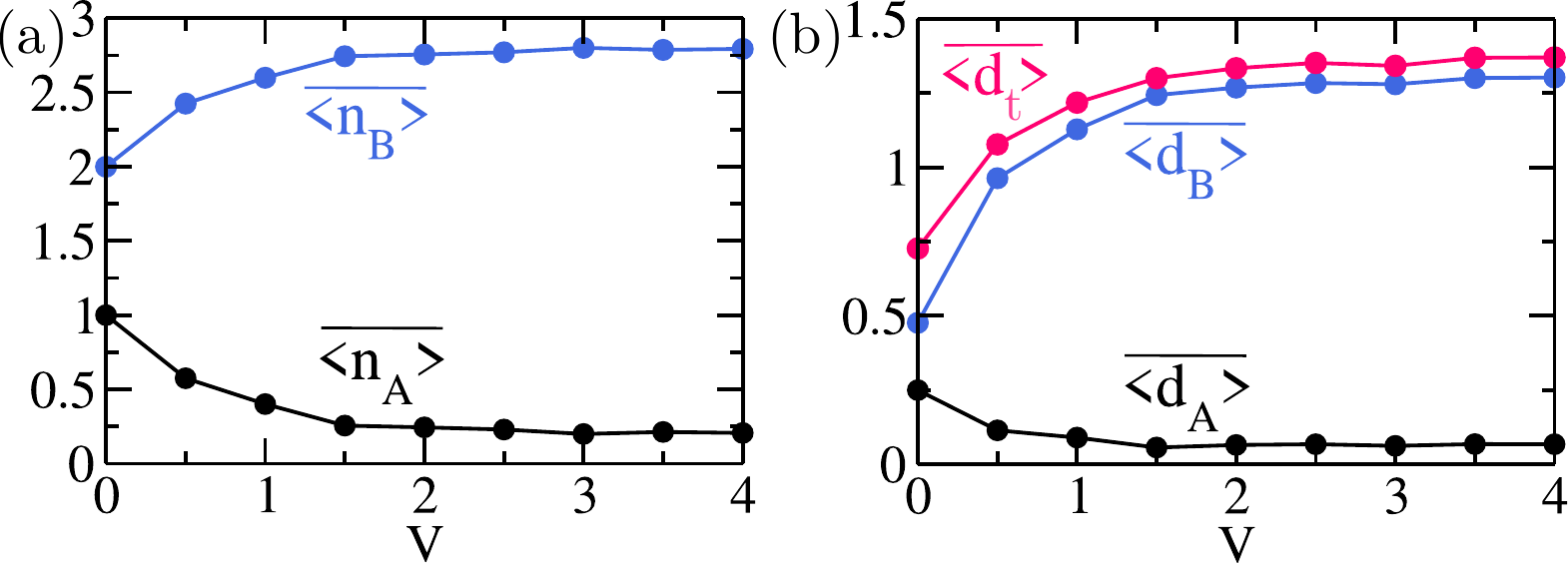}}
\caption{DMRG results for the particle and doublon occupancies for $U=0$ and
$t^\prime=0.8$ as
a function of $V$ in a system of size $L=40$.
(a) Spatial averages of particle occupancy at the $A$ and $B$ sites, $\overline{\braket{n_A}}$ and
$\overline{\braket{n_B}}=\overline{\braket{n_{B_1}}}+\overline{\braket{n_{B_2}}}$, respectively.
(b) Spatial averages of doublon occupancy at $A$ and $B$ sites, $\overline{\braket{d_A}}$ and $\overline{\braket{d_B}}=\overline{\braket{d_{B_1}}}+\overline{\braket{d_{B_2}}}$, respectively,
and at a trimer, $\overline{\braket{d_t}}=\overline{\braket{d_A}}+\overline{\braket{d_B}}$.}
\label{fig:densitiesU0}
\end{figure}

We initially consider the case $U=0$ in Fig. \ref{fig:densitiesU0}, which is
the $t-t^\prime-V$ model for the specified values of $t$ and $t^\prime$.
The average particle and doublon occupancies do not show any sign of
a phase transition at a finite value of $V$. Instead, these particle
distributions suggest that any finite value of $V$ promotes a transfer
of particles from the $A$ sites to the $B$ sites, which is therefore
accompanied by an increase in the average doublon occupancies at the
$B$ sites. In particular, for $V=4$, we find $\overline{\braket{n_A}}\approx 0.25$,
$\overline{\braket{n_B}}=\overline{\braket{n_{B_1}}}+\overline{\braket{n_{B_2}}}\approx 2.75$, and
$\overline{\braket{d_B}}=\overline{\braket{d_{B_1}}}+\overline{\braket{d_{B_2}}}\approx 1.25$.
These values indicate that $V$ induces a coherent superposition
of the states (ii) and (iii) of Fig. \ref{fig:static-conf} and
their respective degeneracy, coexisting with some component
with charge at the sites $A$. Notice, in particular, that for $U=0$, the
states (ii) and (iii) are degenerate in the atomic limit. Thus, the hopping terms
promote a fluctuation between these degenerate configurations
through other states with occupied $A$ sites.
\begin{figure}
\centerline{\includegraphics*[width=0.47\textwidth]{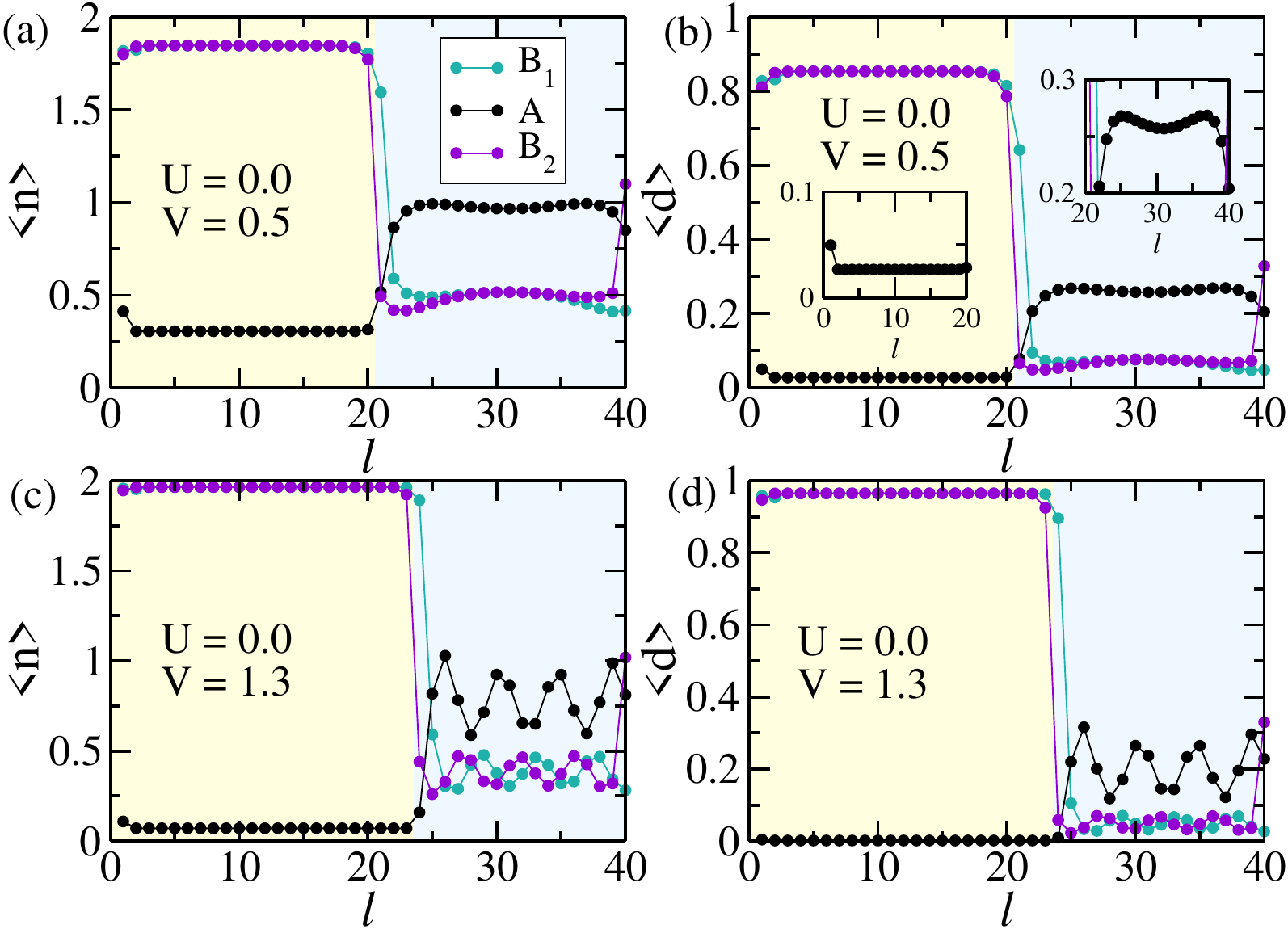}}
\centerline{\includegraphics*[width=0.23\textwidth]{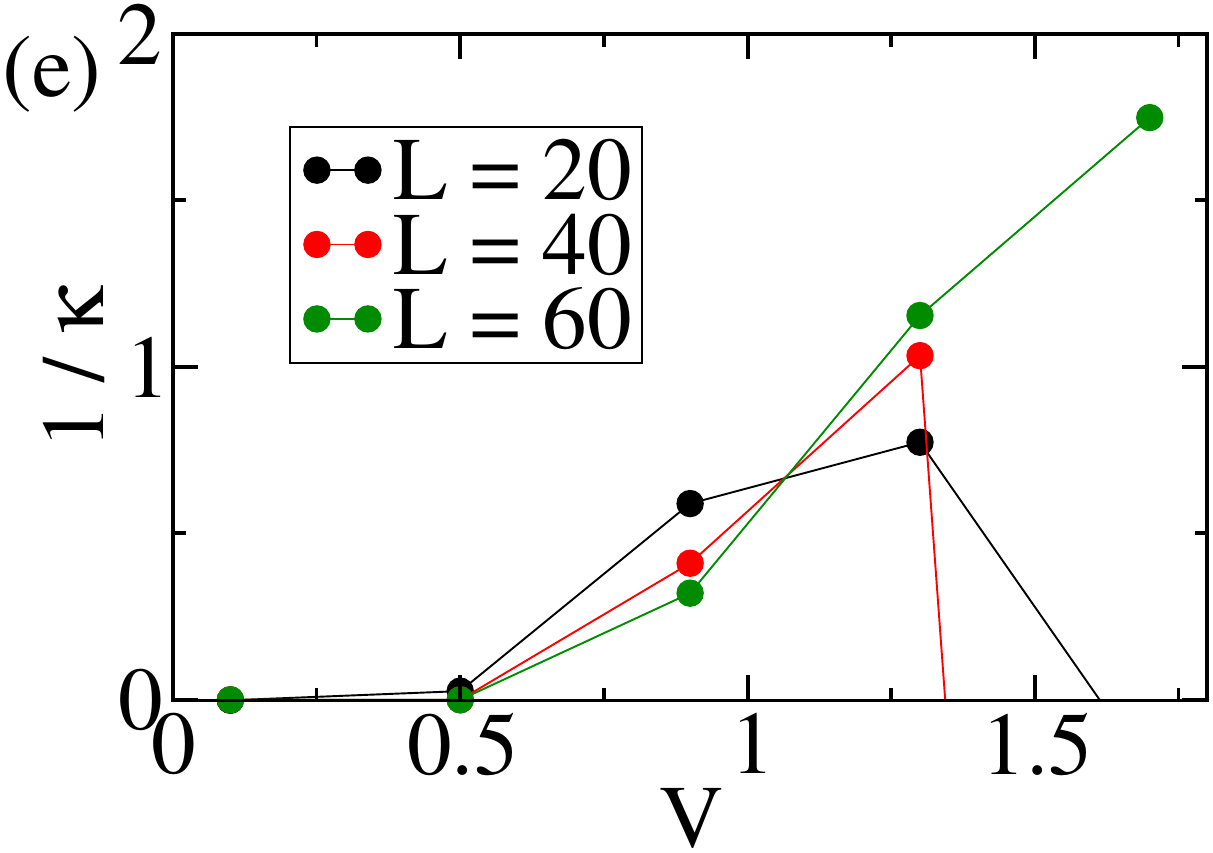}}
\caption{Charge distribution {  and compressibility} in the phase-separation regime for $U= 0$.
DMRG results for the particle and doublon distributions along the chain for $U=0$ and $t^\prime=0.8$ in a system of size $L=40$.
(a) and (c): average particle occupancy $\langle n\rangle$,
(b) and (d): average doublon occupancy $\langle d\rangle$.
In (a) and (b), $V=0.5$; in (c) and (d), { $V=1.3$}. The color code is shown in (a). {The two insets in (b) show a zoomed-in view of the blue and yellow regions of the main figure.} (e) Inverse compressibility $1/\kappa$ as a function
of $V$ for $L=20,40,60$.
}
\label{fig:chargedistU0}
\end{figure}

To characterize the states induced by $V$, we present in Fig. \ref{fig:chargedistU0}
the charge distribution along the trimer lattice for $V=0.5$ and
$V=1.3$. These data show that $V$ leads to phase-separated states in the
lattice, where the two phases have distinct average densities. Thus, for $V\neq 0$,
in the thermodynamic limit, a uniform spatial average density of 3 particles per trimer
is unstable. We notice in Fig. \ref{fig:chargedistU0}
that one of the coexisting phases { (yellow regions)} has trimers occupied by one doublon
at each $B$ site, with a tiny occupancy of the $A$ sites, and that this feature
becomes more prominent for higher values of $V$. This phase has approximately
4 electrons per trimer and is found in {  50\% and 58\% of the lattice for
$V=0.5$ and $V=1.3$, respectively}.
On the other hand, in the second coexisting phase {  (blue regions)},
the main occupancy of the trimers occurs at the $A$ sites, which have a
significant doublon density. In particular, we notice that the doublon occupancy at the $B$
sites is nearly 0 for { $V=1.3$}. {  The second coexisting phase presents a trimer occupancy of
$\approx 2.0$ and $\approx 1.6$ for $V=0.5$ and $V=1.3$, respectively.}

The total spin of the system is null for any finite value of $V$ and $U=0$. {Discarding boundary and interface regions, the uniform average charge distribution in the yellow region suggests an insulating phase, whereas the nonuniform charge distribution in the blue regions indicates a metallic phase.
}

{  In a phase separated regime, the compressibility $\kappa\rightarrow \infty$.
Indeed, we can calculate this quantity to observe its behavior for $U=0$.
The finite-size expression of the compressibility \cite{ogata1991} can be written as
\begin{equation}
 \frac{1}{n^2\kappa}=\frac{N}{4}[E(N_e+2)+E(N_e-2)-2E(N_e)],
 \label{eq:comp}
\end{equation}
where $N_e$ is the total number of electrons, and $n=N_e/N$ is the average particle density, which
in our case is $n=1$. For $V$ greater than 1.0 and $U=0$, the compressibility exhibits
quite erratic behavior, although the state remains phase separated as
exemplified in Figs. \ref{fig:chargedistU0}(c) and \ref{fig:chargedistU0}(d)
for $U=0$ and $V=1.3$.
}

We mention that, due to the specific coexisting phases,
the renormalization procedure typically locks in a metastable state, with
many clusters of the two phases, as exemplified by the charge distributions
in Fig. \ref{fig:chargedistU0}. However, the state of phase separation and
the physical features of the two phases is a confident result from the
numerical data.

\subsection{Complete model: $t\neq 0$, $t^\prime\neq 0$, $U\neq0$, and $V\neq0$}
\begin{figure}
\centerline{\includegraphics*[width=0.5\textwidth]{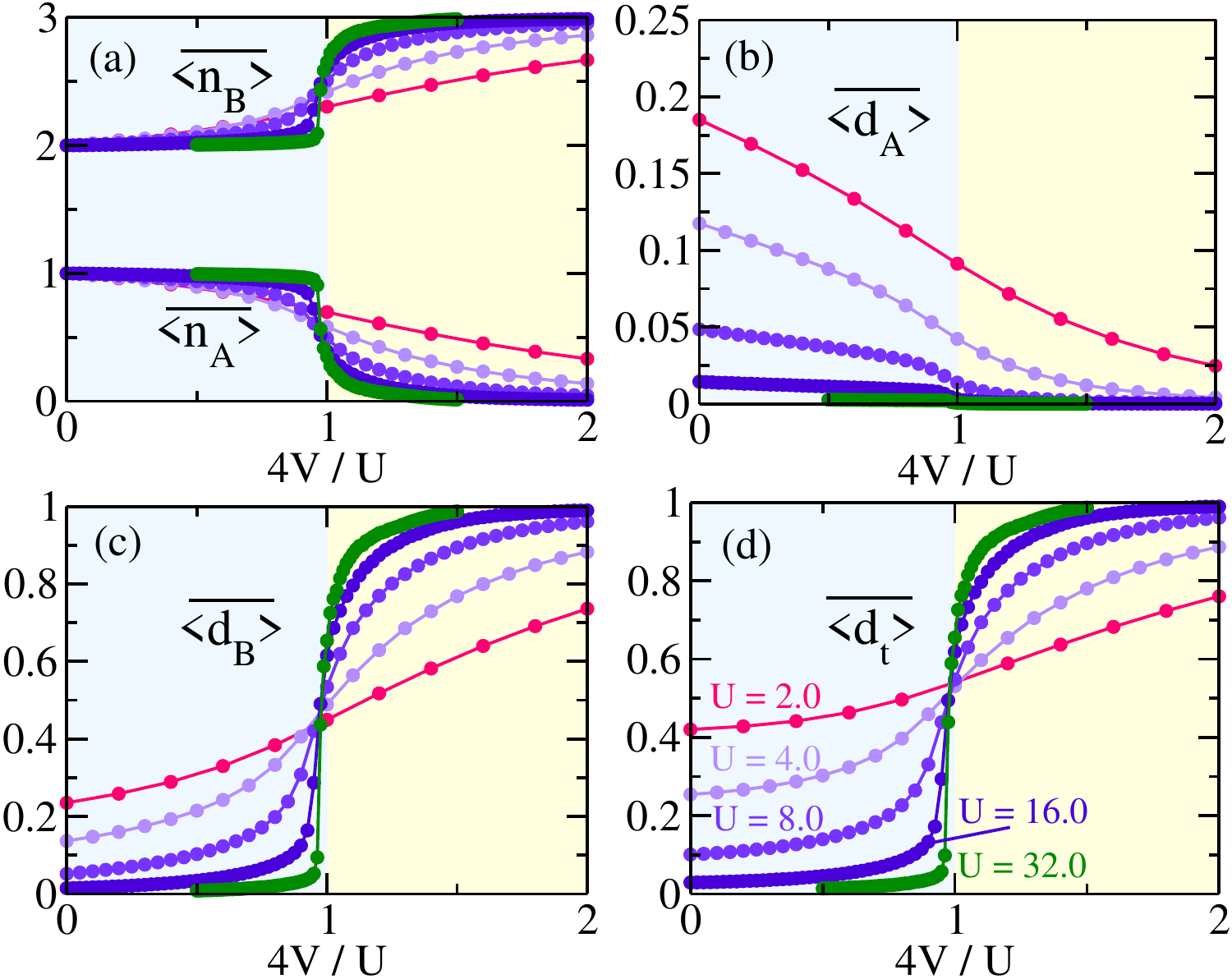}}
\caption{DMRG results for particle and doublon spatial averages for $U\neq 0$ and $t^\prime=0.8$ in a system of size $L=40$.
(a) Spatial average of particle occupancy at $A$ sites, $\overline{\braket{n_A}}$, and at $B$ sites, $\overline{\braket{n_B}}=\overline{\braket{n_{B_1}}}+\overline{\braket{n_{B_2}}}$ for the indicated values of $U$. Spatial average of doublon occupancies: (b) at $A$ sites, $\overline{\braket{d_A}}$, (c) at $B$ sites, $\overline{\braket{d_B}}=\overline{\braket{d_{B_1}}}+\overline{\braket{d_{B_2}}}$, and (d) at a trimer, $\overline{\braket{d_t}}=\overline{\braket{d_A}}+\overline{\braket{d_{B_1}}}+\overline{\braket{d_{B_2}}}$, for the indicated values of $U$.}
\label{fig:chargedistU}
\end{figure}

In Fig. \ref{fig:chargedistU} we display the spatial averages of the particle and doublon occupancies as a function of $V$ for the selected values of $U$, ranging from $U=2.0$ to $U=32.0$. The $V$ axis has been appropriately rescaled by $U/4$, as suggested by Eq. (\ref{eq:trans}). The data delineate a transition that occurs approximately at $V = U/4$. For $V=0$, the mean spatial occupancies of the sites $A$ and $B$ ($B_1$ or $B_2$) correspond to the mean density of 1 electron per site. As in the case $U=0$, the nearest-neighbor Coulomb repulsion $V$ prompts a transfer of charge from the $A$ sites to the two $B$ sites, with the $B$ sites having a spatial average of one doublon and a singly occupied site. However, unlike the scenario where $U=0$, there is an inflection point close to $U/4$ in the curves depicted in Fig. \ref{fig:chargedistU}, with the maximum in the first derivative becoming more pronounced for higher values of $U$. 

\begin{figure}
\centerline{\includegraphics*[width=0.5\textwidth]{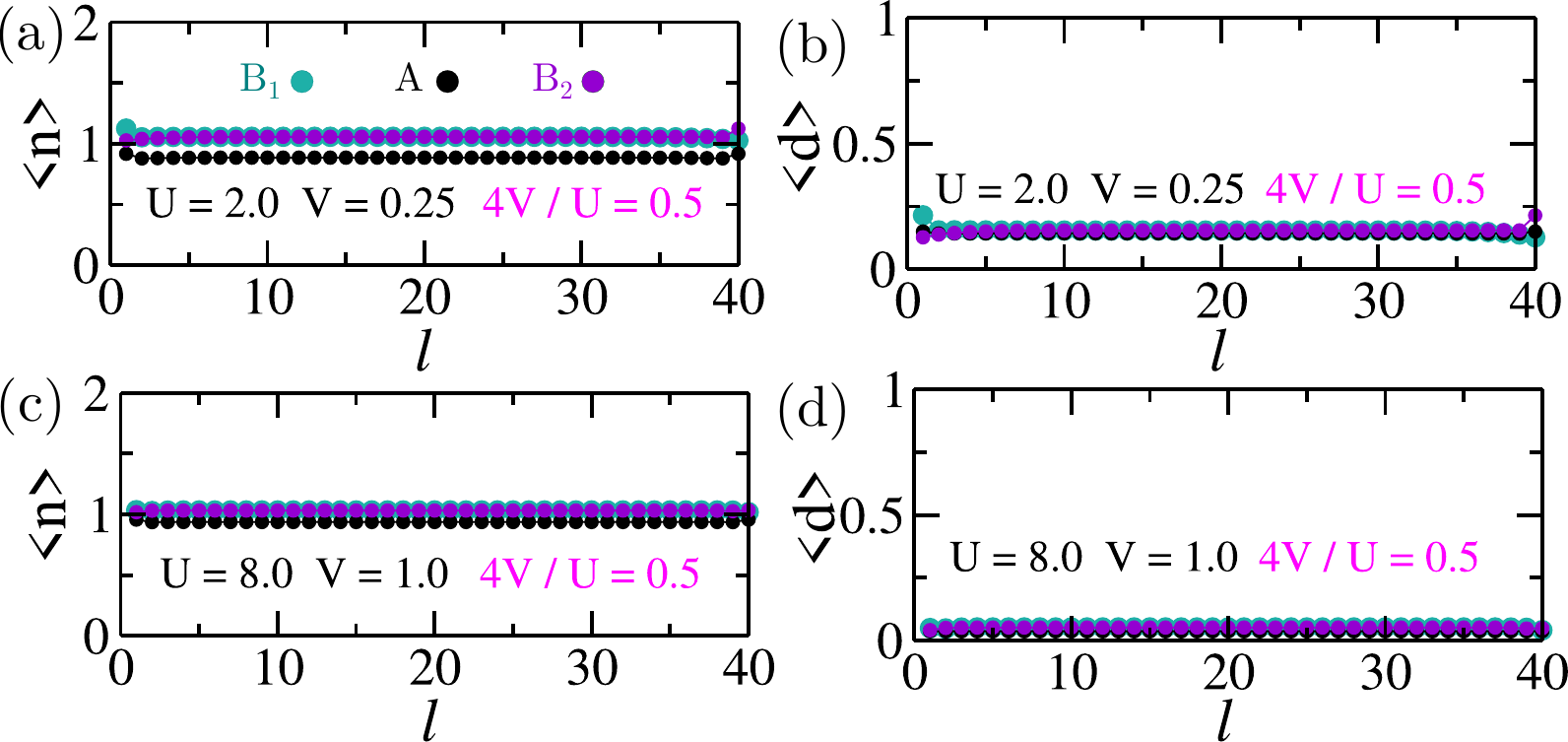}}
\caption{DMRG results for the particle and doublon distributions along the chain for $4V/U=0.5$ and $t^\prime=0.8$ in a system of size $L=40$.  (a) and (c): average particle occupancy $\langle n\rangle$, (b) and (d): average doublon occupancy $\langle d\rangle$. In (a) and (b), $U=2.0$ and $V=0.25$; in (c) and (d), $U=8.0$ and $V=1.0$. The color code is shown in (a).}
\label{fig:localdistUfinitep1}
\end{figure}

We present the average particle and doublon distributions along the chain for $4V/U=0.5$ in Fig. \ref{fig:localdistUfinitep1}. These distributions are associated with a uniform insulating state, with average densities of the $A$ and $B$ sites approximately equal to 1, which is the average particle occupancy of the lattice.  We also notice that the average doublon occupancy of the sites is small for $U=2.0$ and nearly null for $U=8.0$.  
\begin{figure}
\centerline{\includegraphics*[width=0.5\textwidth]{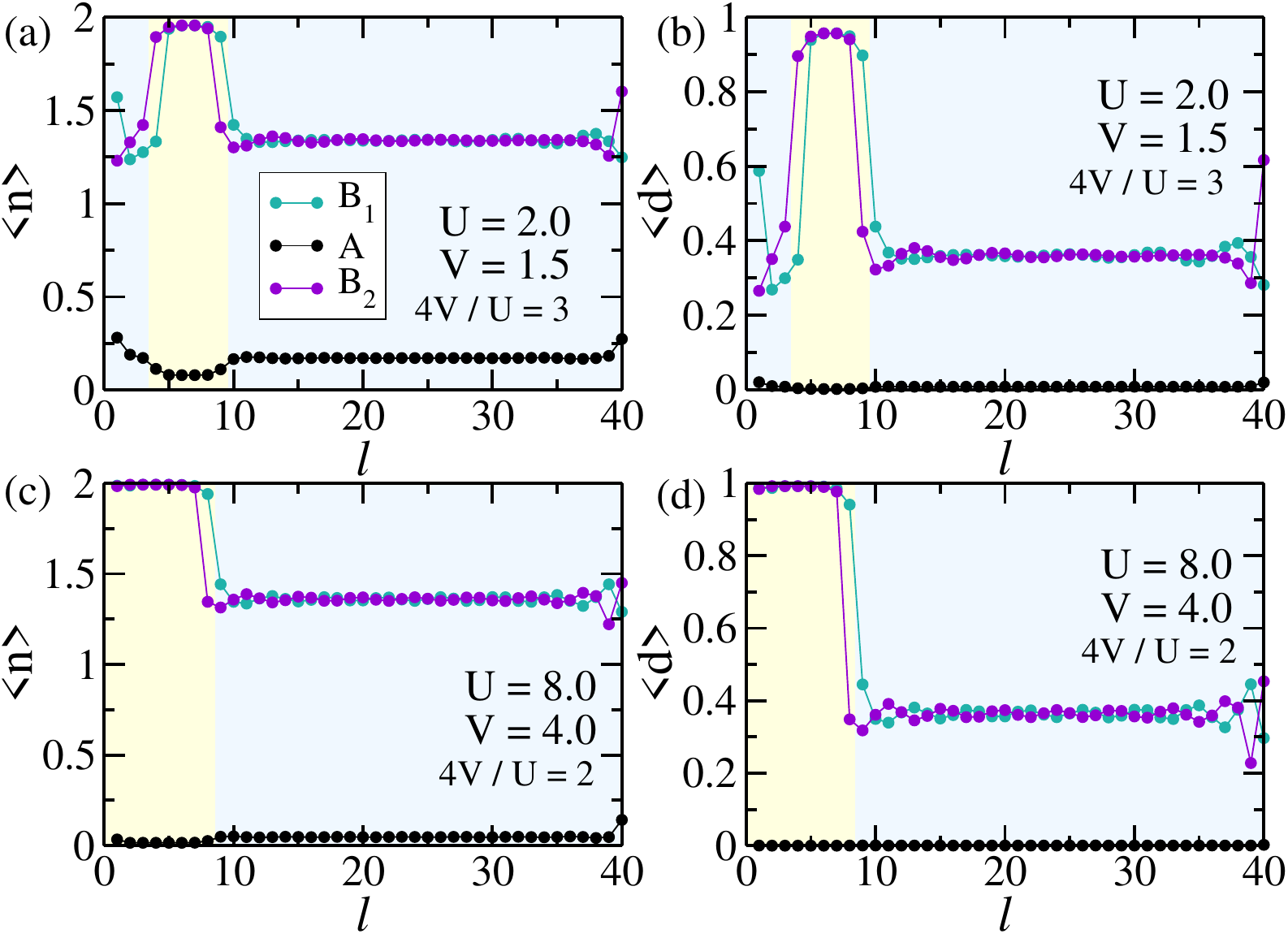}}
\centerline{\includegraphics*[width=0.5\textwidth]{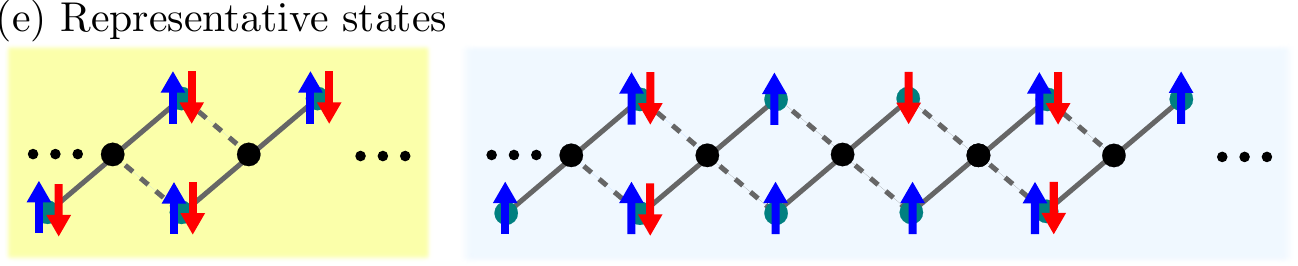}}
\caption{Charge distribution in the phase-separation regime for $U\neq 0$.
DMRG results for the particle and doublon distributions along the chain for $4V/U=2.0$ and 3.0, with $t^\prime=0.8$ in a system of size $L=40$. (a) and (c): average particle occupancy $\langle n\rangle$,  (b) and (d): average doublon occupancy $\langle d\rangle$. In (a) and (b), $U=2.0$ and $V=1.5$; in (c) and (d), $U=8.0$ and $V=4.0$. The color code is shown in (a). The yellow regions highlight phases rich in doublons. (e) Representative states of the yellow and blue regions.}
\label{fig:localdistUfinitep2}
\end{figure}

This scenario changes significantly for $4V/U>1$, as shown in Fig. \ref{fig:localdistUfinitep2}, for which the average density of one particle per site is also unstable. The particle and doublon distributions are nonuniform along the chain, with the coexistence of two spatially separated phases with distinct densities. Although one of the phases{,   yellow regions,} has a high doublon density, as in the case $U=0$ (shown in Fig. \ref{fig:chargedistU0}), the charges essentially occupy the $B$ sites in the {  two coexisting phases}, which was not observed for $U=0$.

The doublon-rich phase, highlighted in yellow in Fig. \ref{fig:localdistUfinitep2}, occupies approximately 18\% of the trimer chain {  for $U=8$ and $V=4$}, with a density of 4 particles per trimer for the values of $U$ shown in Fig. \ref{fig:localdistUfinitep2}. The average particle occupancy of the sites $B_1$ and $B_2$ is $\overline{\braket{n_{B_1}}}=\overline{\braket{n_{B_2}}}\approx 2.0$, while the sites $A$ are weakly populated ($\lesssim 0.05$). Furthermore, in this phase, the average doublon occupancy of the $B_1$ and $B_2$ sites is $\overline{\braket{d_{B_1}}}=\overline{\braket{d_{B_2}}}\approx 1.0$. Thus, a good representative of this phase is the static configuration shown in Fig. \ref{fig:static-conf} (iii) with the substitution of the empty $B$ sites by doubly occupied sites, as sketched in the yellow panel of Fig. \ref{fig:localdistUfinitep2}(e), and quantum fluctuations.

The other phase, highlighted in blue in Fig. \ref{fig:localdistUfinitep2}, occupies nearly 82\% of the lattice {  for $U=8.0$ and $V=4.0$}, and exhibits a density of approximately 2.8 particles per trimer, a decrease from 3 by one particle for every five trimers.  Apart from interface and boundary effects, this decrease matches the excess charge observed in the doublon-rich phase (yellow highlighted regions in Fig. \ref{fig:localdistUfinitep2}). The average particle and doublon occupancies at the $B$ sites are $\overline{\braket{n_{B_1}}}=\overline{\braket{n_{B_2}}}\approx 1.33$, with $\overline{\braket{n_A}}<0.2$, and $\overline{\braket{d_{B_1}}}=\overline{\braket{d_{B_2}}}\approx 0.36$, with the average doublon occupancy of the $A$ sites essentially null. These averages suggest that the configuration in Fig. \ref{fig:static-conf}(ii) with approximately one fewer electron and one fewer doublon for every five trimers, as sketched in the blue panel of Fig. \ref{fig:localdistUfinitep2}(e), giving $\overline{\braket{n_{B_1}}}=\overline{\braket{n_{B_2}}}\approx 1.4$,  $\overline{\braket{d_{B_1}}}=\overline{\braket{d_{B_2}}}\approx 0.4$, and quantum fluctuations represent quite well this phase.

To define the value of $V$ at which the system enters the coexistence region, we consider the quantity $\Delta_{max}$, which is the largest positive deviation of the number of particles in any trimer of the chain from the homogeneous density value of 3 electrons per trimer for fixed values of $U$ and $V$, discarding three trimers at both edges, to minimize boundary effects. For example, in Fig. \ref{fig:localdistUfinitep2}(c), $\Delta_{max}\approx 1$ and is observed at one of the yellow regions.
\begin{figure}
\centerline{\includegraphics*[width=0.5\textwidth]{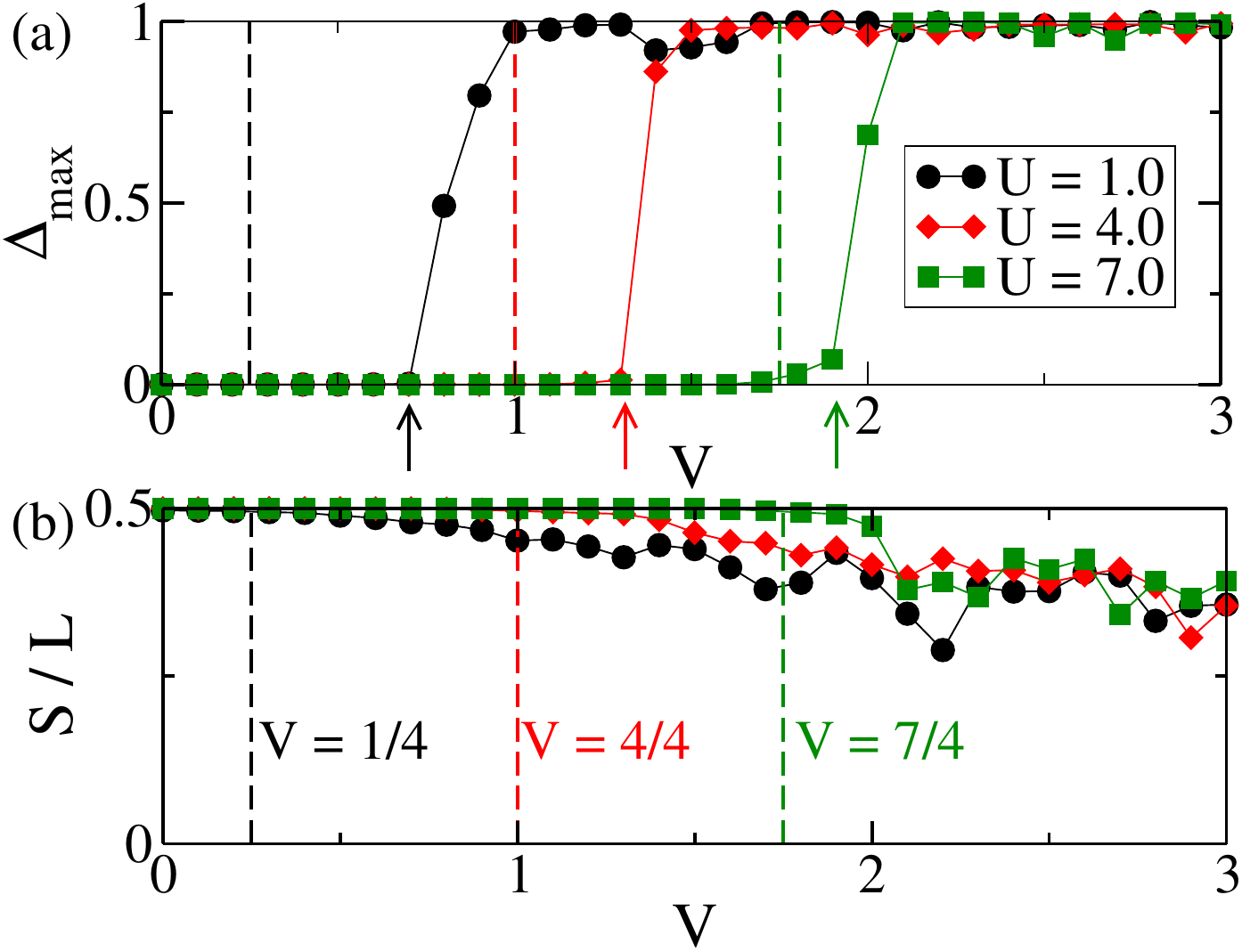}}
\caption{DMRG results for a chain of size $L = 40$ trimers. (a) The largest positive deviation of the density of a trimer, relative to the homogeneous density of 3 electrons per trimer, along the chain, $\Delta_{max}$,  for the indicated values of $U$. 
The arrows indicate the last point where the system presents a homogeneous phase. (b) Spin per trimer, $S/L$, for the indicated values of $U$. Dashed lines indicate the respective approximate crossover $V$ values to the doublon-rich region: $V=U/4$.}
\label{fig:ps}
\end{figure}

In Fig. \ref{fig:ps}(a), we show the behavior of $\Delta_{max}$ as a function of $V$ for three selected values of $U$. The arrows indicate the value of $V$ for which the system is found to be in a homogeneous phase for each value of $U$. In the same figure, we indicate the approximate values of $V$ at which the crossover to the doublon-rich region occurs: $V\approx U/4$, as shown by the data in Fig. \ref{fig:chargedistU}. We observe in Fig. \ref{fig:ps}(a) that there is a range of values of $V$ before phase separation where the doublon density increases significantly (above  $V\approx U/4$), while the system still maintains its homogeneity. However, we notice that the crossover region shrinks as $U$ increases, such that for higher values of $U$ we expect a direct transition to the coexistence region. 

In Fig. \ref{fig:ps}(b), we display the ground-state total spin per trimer, $S/L$, as a function of $V$ for the same values of $U$ shown in Fig. \ref{fig:ps}(a). The value of $S$ is calculated through the spin correlation functions in the sector of the total spin component $S^z=0$ as
\begin{equation}
S(S+1)=\sum_{i,j}\braket{\mathbf{S}_i\cdot\mathbf{S}_j},
\end{equation}
where $\mathbf{S}_i=(S^x_i,S^y_i,S^z_i)$, with $S^\alpha_i$ as the operator of the $\alpha$-component of the spin at site $i$ for $\alpha=x,~y,~z$.  
The Lieb ferrimagnetic phase, with $S/L=1/2$, withstands the increase of $V$, including values of $V$ above the point of crossover to the doublon-rich phase and below the coexistence region. 
Thus, the data show that the Lieb ferrimagnetic phase is robust even in a parameter regime with 
an appreciable number of doublons.
We note that the value of $S/L$ presents some fluctuation in the coexistence region at points for which $S/L\neq 0$. It is due to the presence of nearly degenerate states distinguished in energy by tiny values of interfacial terms between clusters of the two coexisting phases, which have spatial 
particle averages distinct from one particle per site, as shown in Fig. \ref{fig:localdistUfinitep2}. 

{
The phase-separation instability point predicted using $\Delta_{max}$, as shown in Fig. \ref{fig:ps}(a), aligns with the behavior of the compressibility $\kappa$, defined by the finite-size expression in Eq. (\ref{eq:comp}). Figure \ref{fig:comp} displays $\kappa$ as a function of $V$ for the same values of $U$ presented in Fig. \ref{fig:ps}(a), along with the values of $V$ determined by $\Delta_{max}$. In the coexistence region, we observe that $1/\kappa = 0$. Moreover, for $V$ values below the instability point, $1/\kappa$ increases with system size $L$, indicating an incompressible gapped phase: $1/\kappa\rightarrow \infty$ as $L\rightarrow \infty$. This conclusion is further supported by Fig. \ref{fig:comp}(d), which illustrates {$\kappa$ as a function of $1/L$ at the center of the region below the instability point for the specified $U$ values. The fit of the data for the two largest system sizes to the function $\kappa=a/L$, where $a$ is a fitting constant, is excellent, indicating that $\kappa \rightarrow 0$ as $1/L \rightarrow 0$.}

\begin{figure}
\centerline{\includegraphics*[width=0.5\textwidth]{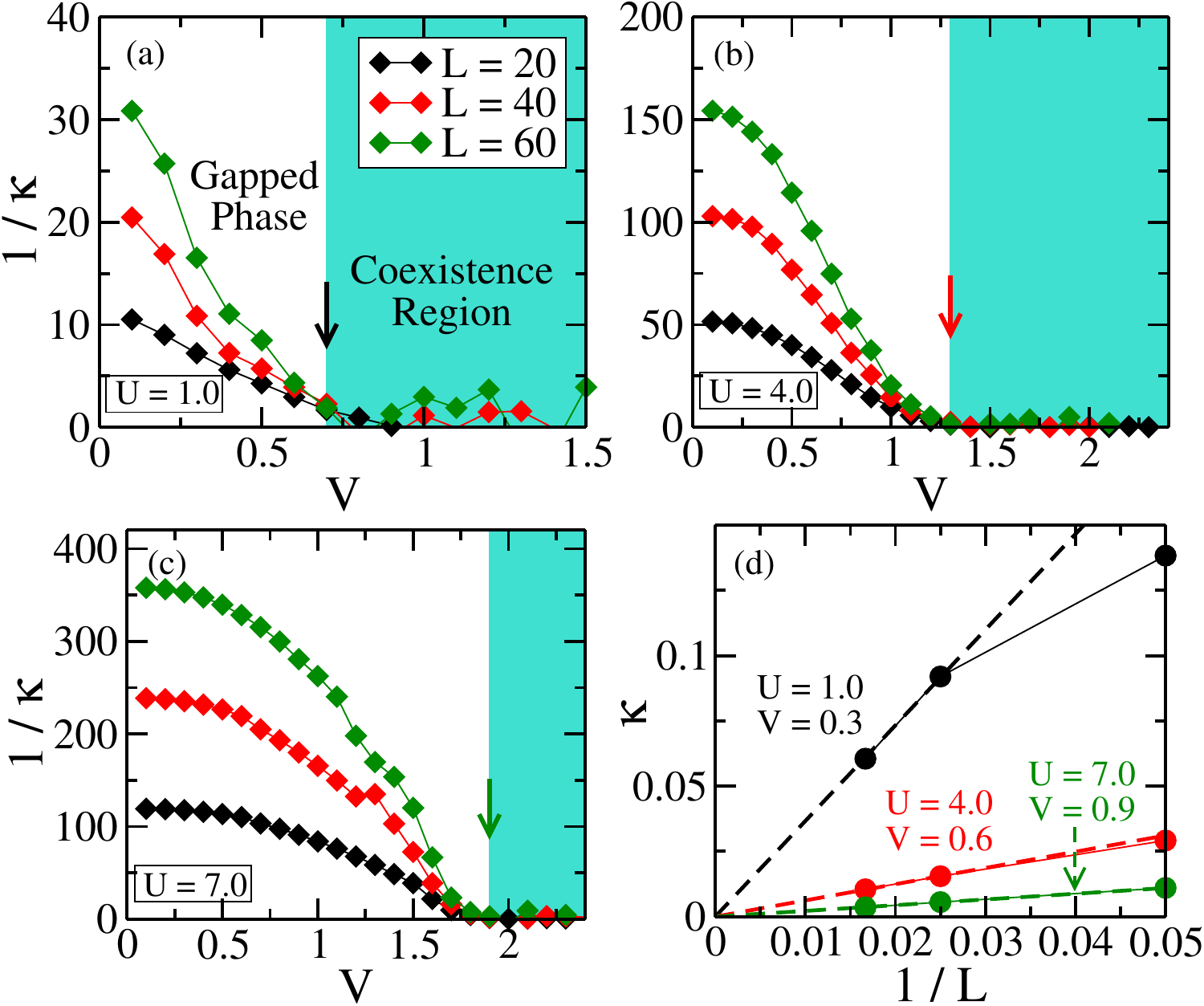}}
\caption{DMRG results for the inverse compressibility $1/\kappa$ for chains of sizes $L = 20,~40,~60$
as a function of $V$ for (a) $U=1.0$, (b) $U=4.0$, and (c) $U=7.0$, which are the same values of $U$ {used} in Fig. \ref{fig:ps}.
The arrows indicate the value of $V$ above which the system becomes unstable, as defined by $\Delta_{max}$, also shown
in Fig. \ref{fig:ps}. {(d) Compressibility $\kappa$ for $U=1.0$ and $V=0.3$, $U=4.0$ and $V=0.6$, $U=7.0$ and $V=0.9$,
as a function of $1/L$. In (d), the dashed lines represent a fit of the data to $\kappa \propto 1/L$ using the two largest system sizes.}}
\label{fig:comp}
\end{figure}

\subsection{Phase diagram}
\begin{figure}
\includegraphics*[width=0.45\textwidth]{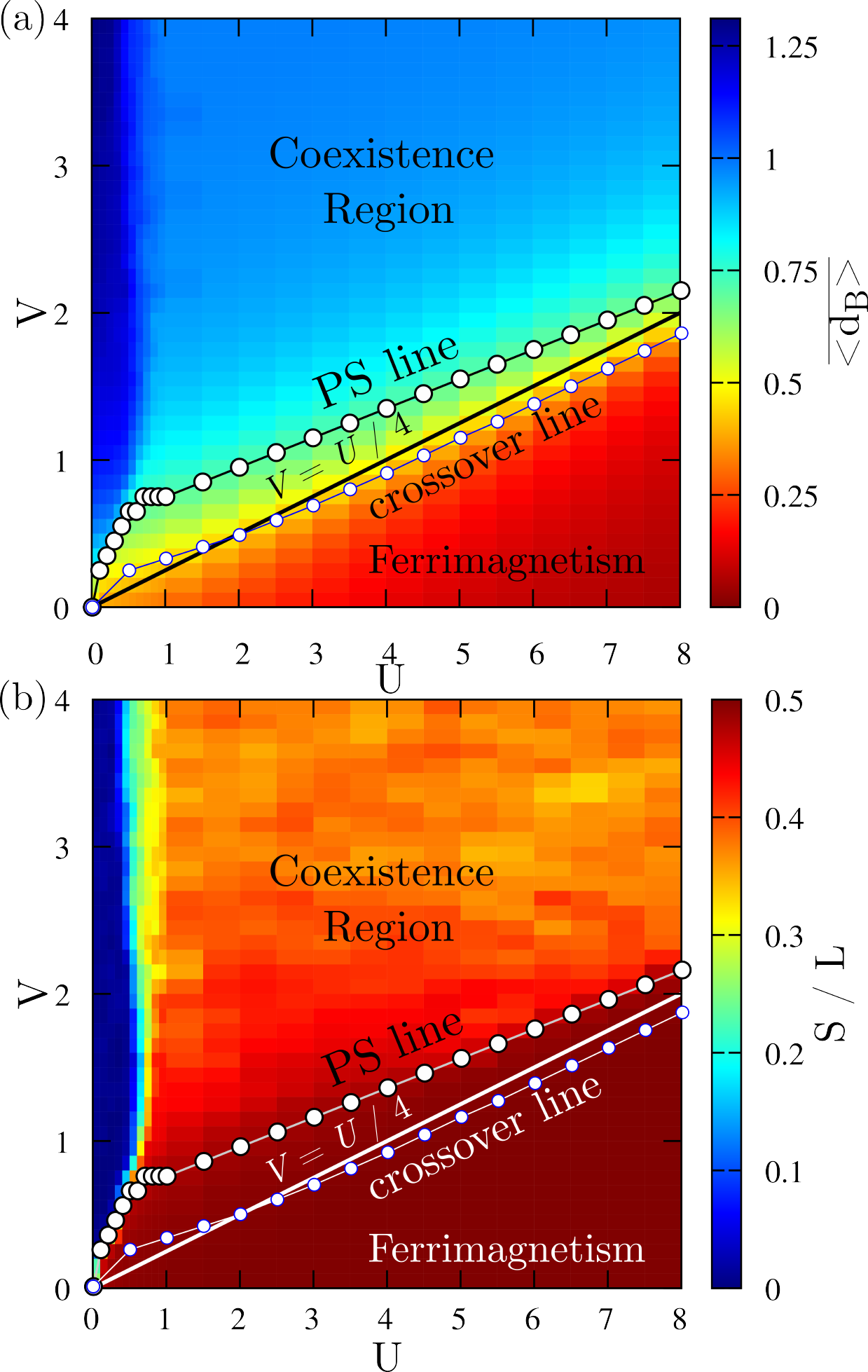}
\caption{DMRG results for (a) the average of doublon occupancy at $B$ sites: $\overline{\braket{d_B}}=\overline{\braket{d_{B_1}}}+\overline{\braket{d_{B_2}}}$,
and (b) total spin per trimer, $S/L$, for a system with $L=40$ trimers in the plane $V$ \textit{versus} $U$. In the Lieb ferrimagnetic phase, $S/L = 1/2$. 
The line $V=U/4$ is a good approximation for the crossover line to the doublon-rich phases. The system enters the coexistence region above the phase-separation (PS) line. Notice that in the blue region of (b), $U\lesssim 1$, the total spin is $\approx 0$ above the PS line. } 
\label{fig:phasediagram}
\end{figure}

We present the spatial average of the doublon density at the $B$ sites, $\overline{\braket{d_B}}=\overline{\braket{d_{B_1}}}+\overline{\braket{d_{B_2}}}$, per trimer, and the spin per trimer in Figs. \ref{fig:phasediagram}(a) and \ref{fig:phasediagram}(b), respectively, for $0\leq V \leq 4$ and $0\leq U \leq 8$, with a sketch of the phase diagram in both figures. The numerical calculation of the inflection point in the curves of $\overline{\braket{d_B}}$ versus $V$ for all available data gives the crossover line, which presents a small departure from $V=U/4$. Below the phase-separation (PS) line, the charge is uniformly distributed along the chain, discarding boundary effects, with a density of 3 electrons per trimer. Above the PS line, we have the coexistence region, in which one of the phases is insulating and has trimers with approximately or exactly 4 electrons at the $B$ sites, as shown in Figs. \ref{fig:chargedistU0}(c) and \ref{fig:localdistUfinitep2}(c).  The other coexisting phase is conducting, with nearly 1 doublon per trimer at the $B$ sites, as shown in Fig. \ref{fig:localdistUfinitep2}(d); and, for $U\lesssim 1$, with the doublons predominantly at the $A$ sites, as shown in Fig. \ref{fig:chargedistU0} (d).

The spin per trimer shown in Fig. \ref{fig:phasediagram}(b) helps to define the magnetic phases of the model. 
We first notice that above the PS line and $U\lesssim 1$, the blue region in Fig. \ref{fig:phasediagram} (b), the coexistence region has $S/L\approx 0$. On the other hand,  the insulating Lieb ferrimagnetic phase, $S/L=1/2$, is robust up to the PS line. In particular, we mention that between the PS line and the crossover line, there is a significant increase in the particle density at the $B$ sites, accompanied by a decrease in the average particle density at the $A$ sites, but maintaining a homogeneous state with a local density of 3 particles per trimer.  Above the PS line, the two coexisting phases do not meet the criterion of one particle per site of Lieb's theorem. However, for $U\gtrsim 1$, in the coexistence region, one of the phases has an unsaturated ferromagnetic moment, with the representative state shown in the blue panel of Fig. \ref{fig:localdistUfinitep2}(e). This conducting phase results from doping the doublon-rich ferrimagnetic state observed between the crossover and PS lines, with some electrons being transferred to the coexisting insulating phase. We propose that the unsaturated ferromagnetic moment arises from a ferrimagnetic state, where the finite magnetic moment on the $A$ sites is antiferromagnetically aligned with the magnetic moment, much higher, of the $B$ sites, and which is supported by the charge itinerancy. In particular, we note that despite the tiny value of the particle occupancy of the $A$ sites, as shown in the blue regions of Fig. \ref{fig:localdistUfinitep2}(c), a coherent fluctuation of the charge in the $B$ sites along this phase, as also observed in those data, requires the charge occupancy of the $A$ sites.  

By calculating the PS line and the crossover line for higher values of $U$, we can estimate the meeting of the two lines at $U\sim 16$. Above this value of $U$, instead of an inflection point, there is a jump in the curves of doublon and particle spatial averages as a function of $V$, signaling the direct transition to the coexistence region without a precedent crossover line and of a doublon-rich Lieb ferrimagnetic phase.

\section{Summary and conclusions}

Using the density matrix renormalization group method, we investigated the extended Hubbard model on a coupled trimer chain at particle density $ n=1 $ (half-filled band). The chain is bipartite, with sublattice sizes $ N_B = 2N_A $, and the intra- and intertrimer hopping amplitudes were set to $ t = 1 $ and $ t' = 0.8 $, respectively. We analyzed the system's behavior as a function of the local on-site Coulomb repulsion $ U $ and the nearest-neighbor interaction $ V $. The phases and their respective transitions were characterized by the spatial averages and distributions of particle and doublon occupancies, along with the total spin per trimer. For $V=0$, the model exhibits a homogeneous ferrimagnetic insulating state with a total spin per trimer $S/L=1/2$, as expected from the Lieb theorem.

For $ U = 0 $ and any value of $ V $, the system separates into two phases, each with an average trimer density distinct from 3 particles per trimer and a total spin per trimer $ S/L = 0 $. One phase is doublon-rich on the $ B $ sublattice, while the other is doublon-rich on the $ A $ sublattice. {  One of the phases is insulating and the other is metallic}.

Moreover, for $ U \neq 0 $, we identified two important lines in the phase diagram: a crossover line and a phase-separation line. The crossover line lies very close to the line $ V = U/4 $, in agreement with the atomic limit of the model. The homogeneous insulating Lieb ferrimagnetic state persists between the crossover and phase separation lines, although a significant charge transfer takes place from the $ A $ to the $B$ sublattice. Furthermore, for $ V $ above the phase-separation line, there exists an inhomogeneous coexistence region where two phases with distinct trimer occupancies coexist in spatially distinct lattice regions. In this coexistence region and $ U \gtrsim 1 $, a ferrimagnetic metallic phase coexists with an insulating paramagnetic phase, characterized by the excess of doublons in sublattice $ B $. Within this phase-separated region, the total spin per trimer satisfies $ S/L \gtrsim 0.3 $, due to the presence of the ferrimagnetic metallic phase. In contrast, for $ U \lesssim 1 $, the coexistence region resembles that of $ U = 0 $ {  case, with two paramagnetic phases coexisting}.

Our results will certainly stimulate future relevant investigations, including the doped model away from half-filling, topological aspects of the underlying phase transitions, and the nature of the edge states in open chains.
 
\begin{acknowledgments}

We acknowledge the support from the Brazilian agencies Coordenação de Aperfeiçoamento de Pessoal de Nível Superior (CAPES), Grants No. PROEX 534/2018, No. 23038.003382/2018-39; and Fundação de Amparo à Ciência e Tecnologia do Estado de Pernambuco (FACEPE), research grant PRONEX Program funded by CNPq and FACEPE, APQ-0602-1.05/14. M.D.C.-F. acknowledges the fellowship grant No. 304646/2019-9 from Conselho Nacional de Desenvolvimento Científico e Tecnológico (CNPq).
\end{acknowledgments}

\end{document}